\documentclass[11pt,a4paper]{article}
\pdfoutput=1

\usepackage{jcappub}
\usepackage[utf8]{inputenc}
\usepackage{graphicx}
\usepackage{epsfig}
\usepackage{subfigure}
\usepackage{color}
\usepackage{comment}
\usepackage{amsmath}
\usepackage{mathtools}
\usepackage{tabularx}
\newcommand{\be}{\begin{equation}} % only untightened
\newcommand{\ee}{\end{equation}}
\newcommand{\bea}{\begin{equation}\begin{aligned}} 
\newcommand{\eea}{\end{aligned}\end{equation}}

\newcommand{\bmp}{\noindent\begin{minipage}{16cm}}
\newcommand{\emp}{\end{minipage}\vskip 7mm} % 7mm untightened
\def\lsim{\mathrel{\raise.3ex\hbox{$<$\kern-.75em\lower1ex\hbox{$\sim$}}}}
\def\gsim{\mathrel{\raise.3ex\hbox{$>$\kern-.75em\lower1ex\hbox{$\sim$}}}}

\newcommand{\intron}[1]{}%{#1}

\providecommand{\f}[2]{\frac{{#1}}{{#2}}}

\newcommand{\dd}{\mathrm{d}}

\title{Scalar correlation functions in de~Sitter space from the stochastic {spectral expansion}}

\author[a,b]{Tommi Markkanen}
\author[a]{, Arttu Rajantie}
\author[c]{, Stephen Stopyra}
\author[d]{and Tommi Tenkanen}

\affiliation[a]{Department of Physics, Imperial College London,\\Blackett Laboratory, London, SW7 2AZ, United Kingdom}
\affiliation[b]{Laboratory of High Energy and Computational Physics, National Institute of Chemical Physics and
Biophysics, R\"avala pst. 10, Tallinn, 10143, Estonia}
\affiliation[c]{Department of Physics and Astronomy, UCL, Gower Street, London WC1E 6BT, United Kingdom}
\affiliation[d]{Department of Physics and Astronomy, Johns Hopkins University, \\
Baltimore, MD 21218, United States of America}
    
\emailAdd{t.markkanen@imperial.ac.uk}
\emailAdd{tommi.markkanen@kbfi.ee}
\emailAdd{a.rajantie@imperial.ac.uk}
\emailAdd{s.stopyra@ucl.ac.uk}
\emailAdd{ttenkan1@jhu.edu}

\abstract{We consider light scalar fields during inflation and show how the {stochastic} spectral expansion method can be used to calculate two-point correlation functions of an arbitrary local function of the field in de Sitter space. In particular, we use this approach for a massive scalar field with quartic self-interactions to calculate the fluctuation spectrum of the {density contrast} and compare it to other approximations. We find that neither Gaussian nor linear approximations accurately reproduce the power spectrum, {and in fact always overestimate it}. {For example,} for a scalar field with only a quartic term in the potential, $V=\lambda\phi^4/4$, we find a blue spectrum with spectral index $n-1=0.579\sqrt{\lambda}$.}

\begin{document}
\begin{flushleft}
	\hfill		  IMPERIAL/TP/2019/TM/03 \\
\end{flushleft}
\maketitle
%%%%%%%%%%%%%%%%%%%%%%%%%%%%%%%%%%%%%%%%%%%%%%%%%%%%%%%%%%%%%%%%%%%%%%%%%%%%%%%%%%%%%%%%%%%%%%%%%%%%

%%%%%%%%%%%%%%%%%%%%%%%%%%%%%%%%%%%%%%%%%%%%%%%%%%%%%%%%%%%%%%%%%%%%%%%%%%%%%%%%%%%%%%%%%%%%%%%

\section{Introduction}

Light scalar spectator fields acquire super-horizon fluctuations during cosmological inflation with potentially observable consequences. For example, they can give rise to curvature perturbations through the curvaton mechanism~\cite{Linde:1996gt,Enqvist:2001zp,Lyth:2001nq,Moroi:2001ct} or through their effect on the non-equilibrium reheating dynamics at the end of inflation~\cite{Enqvist:2004ey,Chambers:2007se}. They may also influence the dark matter abundance~\cite{Peebles:1999fz,Nurmi:2015ema,Markkanen:2018gcw}, the anisotropy of the gravitational wave background~\cite{Bethke:2013aba}, generate primordial black holes~\cite{Kawasaki:2012wr,Kohri:2012yw,Carr:2017edp,Carr:2019hud} {or} source Dark Energy \cite{Glavan:2017jye}.
{In particular,} the inflationary dynamics of the Standard Model Higgs~\cite{Enqvist:2014tta,Enqvist:2015sua,Figueroa:2015rqa} have a wide variety of possible ramifications, from gravitational waves~\cite{Figueroa:2014aya,Figueroa:2016ojl} and leptogenesis~\cite{Kusenko:2014lra,Pearce:2015nga} to triggering electroweak vacuum decay~\cite{Espinosa:2007qp,Herranen:2014cua,Markkanen:2018pdo}.

In the case of free fields the scalar field fluctuations can be computed by solving the mode equations on a curved background~\cite{Birrell:1982ix}. However, when self-interactions are important it is well-known that for light fields the perturbative expansion fails~\cite{Allen:1985ux,Allen:1987tz}, requiring one to employ {other techniques.} %resummations. 
The {\em stochastic %Starobinsky-Yokoyama
approach}~\cite{Starobinsky:1986fx,Starobinsky:1994bd} provides a powerful tool for such computations and is in agreement with, 
{and} in fact often superior to, other approaches to quantum field theory in de Sitter space~\cite{Hu:1986cv,Boyanovsky:2005sh,Tsamis:2005hd,Serreau:2011fu,Tokuda:2017fdh,Arai:2011dd,Guilleux:2015pma,Prokopec:2017vxx,Moss:2016uix}. In particular we note the direct QFT approaches making use of the Schwinger-Dyson equations: in Ref. \cite{Gautier:2013aoa} to two-loop order in perturbation theory and in Ref. \cite{Gautier:2015pca} to Next-To-Leading order in a $1/N$ expansion, where the latter is also in agreement with the recent Euclidean analysis of Ref. \cite{LopezNacir:2019ord}.

In many inflationary models, the spacetime is well approximated by a de Sitter space and inflation lasts for sufficiently long for the field fluctuations to equilibrate \cite{Enqvist:2012xn,Grain:2017dqa}. In that case, the problem of finding scalar correlators reduces to finding the equilibrium state. The %Starobinsky-Yokoyama 
stochastic approach gives an exact expression for the one-point equilibrium probability distribution of the field, from which a spectral expansion for two-point correlation functions can be obtained~\cite{Starobinsky:1994bd}.

The one-point probability distribution is well-known and widely used, see for example {Ref.}~\cite{Hardwick:2017fjo}. Investigating correlators via the stochastic formalism has been less common, however  see~\cite{Peebles:1999fz,Tsamis:2005hd,Prokopec:2007ak,Riotto:2008mv,Kunimitsu:2012xx,Motohashi:2012bb,Garbrecht:2013coa,Garbrecht:2014dca,Burgess:2014eoa,Onemli:2015pma,Prokopec:2015owa,Cho:2015pwa,Vennin:2015hra,Kitamoto:2018dek,Markkanen:2018gcw},
even though it is the correlators that are more often directly related to observations. In particular the isocurvature perturbations, which are heavily constrained by observations~\cite{Aghanim:2018eyx}, were studied in  {Refs.}~\cite{Beltran:2006sq,Kainulainen:2016vzv,Heikinheimo:2016yds,Enqvist:2017kzh,Graham:2018jyp,Markkanen:2018gcw} with the stochastic approach. %(see also Ref. \cite{Guth:2018hsa}).

The aim of this paper is to use the stochastic approach to compute scalar field correlators in de Sitter space for a scalar field $\phi$ with a mass term $\f12 m^2\phi^2$ and a quartic self-interaction term $\f\lambda4 \phi^4$. Up to trivial scaling, these are parameterised by a single dimensionless parameter $\alpha\equiv{m^2}/({\sqrt{\lambda}H^2})$. The method involves finding eigenfunctions and eigenvalues of a Schr\"odinger-like equation, which is easy to do numerically to high precision, and which give the coefficients and exponents for an asymptotic long-distance expansion of correlators. 
We compute these numerically for $0\le \alpha\le \infty$. We also consider perturbative expansions around the two limits $\alpha\rightarrow 0$ and $\alpha\rightarrow\infty$, which correspond to the massless and free cases, which have been studied in the literature earlier. Comparison of the results shows that the commonly used {mean field} % linear 
and Gaussian approximations do not describe the field correlators correctly away from the free limit.

The paper is organised as follows: in Section \ref{dynamics}, we summarise the stochastic approach and discuss one- and two-point correlators of local functions of a generic light scalar field in de Sitter space. In Section \ref{sec:examples}, we discuss as an example the case of a massive self-interacting field, computing the relevant eigenvalues and eigenfunctions related to the spectral expansion of correlators. In Section \ref{sec:comparison}, 
we compare the results with different approximations. Finally, in Section \ref{sec:conclusions}, we conclude with an outlook. 

\section{The stochastic approach}
\label{dynamics}

\subsection{One-point probability distribution}
It was shown by Starobinsky and Yokoyama~\cite{Starobinsky:1986fx,Starobinsky:1994bd} that on super-horizon scales, the behaviour of a light and energetically subdominant scalar field $\phi$ in de Sitter spacetime is described by a stochastic Langevin equation
\begin{equation}
\dot\phi(t)=-\frac{1}{3H}V'(\phi)+\xi(t),
\label{eq:Langevin}
\end{equation}
where $H=\dot{a}/a$ is the Hubble rate, $V(\phi)$ is the potential of the scalar field and prime denotes derivative with respect to the field, and $\xi(t)$ is a Gaussian white noise term with two-point correlator
\begin{equation}
\langle \xi(t_1) \xi(t_2)\rangle = \frac{H^3}{4\pi^2}\delta(t_1-t_2).
\end{equation}
The condition for the field being light is $V''(\phi)<H$, whereas the field is energetically subdominant when $V<3H^2M_{\rm P}^2$, where $M_{\rm P}$ is the reduced Planck mass.

Let us denote the one-point probability distribution of the scalar field $\phi$ at time $t$ by $P(t;\phi)$. Starobinsky and Yokoyama showed \cite{Starobinsky:1994bd} that it satisfies the Fokker-Planck equation
\begin{equation}
\frac{\partial P(t;\phi)}{\partial t}=D_\phi P(t;\phi),\label{eq:FP}
\end{equation}
where $D_\phi$ is the differential operator
\begin{equation}
D_\phi\equiv
\frac{V''(\phi)}{3H}
+
\frac{V'(\phi)}{3H}
\frac{\partial}{\partial\phi}
+\frac{H^3}{8\pi^2}\frac{\partial^2}{\partial \phi^2}
.
\end{equation}
From this it is straightforward to find the equilibrium distribution
\be
P_{\rm eq}(\phi)\propto \exp\left(-\frac{8\pi^2V(\phi)}{3H^4}\right),
\label{eq:eqP}
\ee
up to a normalization factor which guarantees total probability equal to unity.

If the potential is symmetric under $\phi\rightarrow -\phi$, the average value of the field vanishes,
\be
\langle\phi\rangle=0,
\ee
but in any given Hubble volume the field is non-zero.
Note that because of the non-Gaussian form of the equilibrium distribution and its zero mean, it would be incorrect to use results that assume either Gaussianity or linear fluctuations around a non-zero mean {field} value.

Considering, as an example, a simple quartic potential,
\be
V(\phi)=\frac{1}{4}\lambda\phi^4,
\ee
one finds the field variance~\cite{Starobinsky:1994bd}
\be 
\label{hstar}
\langle \phi^2\rangle = \left(\frac{3}{2\pi^2}\right)^{1/2}{\frac{\Gamma(\frac{3}{4})}{\Gamma(\frac{1}{4})}} \frac{H^2}{\lambda^{1/2}},
\ee 
and the average potential energy of  the field~\cite{Starobinsky:1994bd}
\be
\langle V(\phi)\rangle=\frac{3H^4}{32\pi^2}.\label{eq:potV}
\ee
%For comparison, if we were to (incorrectly) assume that the distribution is Gaussian we could write
%\be
%\langle V(\phi)\rangle_{\rm Gauss}=\frac{3}{4}\lambda\langle \phi^2\rangle^2
%=\frac{9}{8\pi^2}\left(\frac{\Gamma(\frac{3}{4})}{\Gamma(\frac{1}{4})}\right)^2H^4
%\approx 1.371
%\rho.
%\ee
%Therefore the Gaussian approximation gets the answer wrong by approximately 37\%.

\subsection{Temporal correlation functions}

Let us now move to consider two-point correlation functions. 
We are interested in the general two-point function of some local function $f(\phi)$ of the field,
\be
G_f(t_1,t_2;\vec{r}_1,\vec{r}_2)\equiv
\langle f(\phi(t_1,\vec{r}_1) f(\phi(t_2,\vec{r}_2)\rangle.
\ee
Starobinsky and Yokoyama developed a method for calculating them~\cite{Starobinsky:1994bd}, but it has been used much less than the one-point probability distribution~(\ref{eq:eqP}). In the following we present their calculation.

First, note that because of de Sitter invariance~\cite{Starobinsky:1994bd}, any correlator of a scalar observable $f(\phi)$ can only depend on the de Sitter invariant quantity
%(Note: Here $\\vec{x}$ is comoving...)
\be
y=\cosh H(t_1-t_2)-\frac{H^2}{2}e^{H(t_1+t_2)}|\vec{r}_1-\vec{r}_2|^2,
\ee
where $\vec{r}_{1}$ and $\vec{r}_2$ are comoving position vectors.
As long as $|y|\gg 1$, both time-like and space-like separations can be expressed as
\be
G_f(t_1,t_2;\vec{r}_1,\vec{r}_2)=G_f\left(H^{-1}\ln\left|2y-1\right|\right),
\ee
where the right-hand-side is the temporal correlation function
\be
G_f(t)\equiv G_f(t;0)
=\langle f(\phi(0)) f(\phi(t))\rangle
,
\label{equ:Gfdef}
\ee
corresponding to $\vec{r}_2=\vec{r}_1$,
which can be obtained easily from the stochastic approach.

Following Starobinsky and Yokoyama~\cite{Starobinsky:1994bd}, let us  define
\begin{equation}
\tilde{P}(t;\phi)=e^\frac{4\pi^2 V(\phi)}{3H^4}P(t;\phi),
\end{equation}
which satisfies the equation
\begin{equation}
\frac{\partial \tilde{P}(t;\phi)}{\partial t}=\frac{3H^3}{4\pi^2}\tilde{D}_\phi \tilde{P}(t;\phi),\label{eq:egs}
\end{equation}
with
\begin{equation}
\tilde{D}_\phi=\frac{1}{2}\frac{\partial^2}{\partial\phi^2}-\frac{1}{2}\left(v'(\phi)^2-v''(\phi)\right),
\quad
v(\phi)=\frac{4\pi^2}{3H^4}V(\phi).\label{eq:defs}
\end{equation}

Because this is a linear equation, we can use separation of variables to 
find independent solutions of the form
$\tilde{P}_n(t;\phi)=e^{-\Lambda_n t}\psi_n(\phi)$,
where $\psi_n(\phi)$ satisfies the time-independent Schr\"odinger-like eigenvalue equation
\begin{equation}
\tilde{D}_\phi\psi_n(\phi)
=-\frac{4\pi^2\Lambda_n}{H^3}\psi_n(\phi).\label{e:sch}
\end{equation} 
We assume that the eigenfunctions $\psi_n(\phi)$ are orthonormal,
\begin{equation}
\int d\phi \psi_m(\phi)\psi_n(\phi)=\delta_{m,n},
\end{equation}
and complete
\begin{equation}
\sum_n \psi_n(\phi)\psi_n(\phi_0)=\delta(\phi-\phi_0).
\end{equation}

The lowest eigenvalue, with $\Lambda_0=0$,
is
\begin{equation}
\psi_0(\phi)\propto e^{-\frac{4\pi^2V(\phi)}{3H^4}},\label{eq:p0}
\end{equation}
and comparing with Eq.~(\ref{eq:eqP}), one can see that
\begin{equation}
P_{\rm eq}(\phi)=\psi_0(\phi)^2.
\label{equ:Peq}
\end{equation}
It is convenient to introduce Dirac-like notation
\be
\langle n|f|m\rangle\equiv\int d\phi \psi_n(\phi)f(\phi)\psi_m(\phi).
\ee
In this way we can, for example, write local equilibrium expectation values as
\be
\langle f(\phi)\rangle = 
\int d\phi \psi_0(\phi)f(\phi)\psi_0(\phi)=\langle 0 | f | 0 \rangle.
\ee

Because of the linearity, the probability distributions 
between any two times are related linearly,
\begin{equation}
\tilde{P}(t+\Delta t;\phi)=\int d\phi_0 \tilde{U}(\Delta t;\phi,\phi_0)\tilde{P}(t;\phi_0),
\end{equation}
by a transfer matrix $\tilde{U}(\Delta t;\phi,\phi_0)$,
which satisfies the same differential equation
\begin{equation}
\frac{\partial \tilde{U}(t;\phi,\phi_0)}{\partial t}=\frac{3H^3}{4\pi^2}\tilde{D}_\phi \tilde{U}(t;\phi,\phi_0).
\end{equation}
Using this we can also write for the unscaled probability distribution
\begin{eqnarray}
P(t+\Delta t;\phi)&=&
\int d\phi_0 e^\frac{-4\pi^2V(\phi)}{3H^4}\tilde{U}(\Delta t;\phi,\phi_0)
e^\frac{4\pi^2V(\phi_0)}{3H^4}
P(t;\phi_0)
\nonumber\\
&=&
\int d\phi_0 U(\Delta t;\phi,\phi_0)
P(t;\phi_0)
,
\end{eqnarray}
where we have defined
\begin{equation}
U(\Delta t;\phi,\phi_0)\equiv
e^\frac{-4\pi^2V(\phi)}{3H^4}\tilde{U}(\Delta t;\phi,\phi_0)
e^\frac{4\pi^2V(\phi_0)}{3H^4}.
\end{equation}
The transfer matrix satisfies the initial condition 
\begin{equation}
U(0;\phi,\phi_0)=\tilde{U}(0;\phi,\phi_0)=\delta(\phi-\phi_0),
\end{equation}
and can be written in terms of the eigenfunctions as
\begin{equation}
\tilde{U}(t;\phi,\phi_0)=\sum_n e^{-\Lambda_nt}\psi_n(\phi)\psi_n(\phi_0).
\label{equ:U_spectral}
\end{equation}

To obtain the temporal correlation function {(\ref{equ:Gfdef})}, %$G_f(t)$, 
we integrate over $\phi_0$ weighted by the equilibrium distribution $P_{\rm eq}(\phi_0)$,
\begin{eqnarray}
G_f(t)
&=&
\int d\phi\int d\phi_0 P_{\rm eq}(\phi_0) f(\phi_0) U(t;\phi,\phi_0)f(\phi)
\nonumber\\
&=&\int d\phi\int d\phi_0 P_{\rm eq}(\phi_0) f(\phi_0) e^{\frac{4\pi^2}{3H^4}(V(\phi_0)-V(\phi))}
\tilde{U}(t;\phi,\phi_0)f(\phi)
.
\label{equ:2point}
\end{eqnarray}

Substituting Eq.~(\ref{equ:Peq}) into Eq.~(\ref{equ:2point}), we find
\begin{eqnarray}
{G_f(t)}&=&
\int d\phi\int d\phi_0  \psi_0(\phi_0)f(\phi_0) 
\tilde{U}(t;\phi,\phi_0)\psi_0(\phi) f(\phi)
.
\end{eqnarray}
Now, we substitute the spectral expansion (\ref{equ:U_spectral})
\begin{eqnarray}
{G_f(t)}&=&\sum_n \langle 0 | f |n \rangle 
e^{-\Lambda_nt}
\langle n | f |0 \rangle
=
\sum_nf_n^2 e^{-\Lambda_nt}
,
\label{equ:spectral}
\end{eqnarray}
where
\begin{equation}
f_n=\langle 0 | f | n \rangle=\int d\phi \psi_0(\phi)f(\phi)\psi_n(\phi)\label{eq:fn}
.
\end{equation}
Therefore the asymptotic behaviour is given by the lowest eigenvalues $\Lambda_n$.

If the potential $V(\phi)$
is symmetric under parity $\phi\rightarrow -\phi$,
the eigenfunctions are either odd or even. The lowest eigenfunction $\psi_0$ is even, so if we are considering an even/odd function $f(\phi)$, only the even/odd eigenvalues contribute, respectively.
For example, for the field correlator $f(\phi)=\phi$, the lowest contributing eigenvalue is 
$\Lambda_1$. This gives
\begin{equation}
\langle \phi(0)\phi(t)\rangle 
{=G_\phi(t)} = \phi_1^2e^{-\Lambda_1t} + \mathcal{O}(e^{-\Lambda_3t}),
\end{equation}
where 
\begin{equation}
\phi_1=\int d\phi \psi_0(\phi)\phi \psi_1(\phi)
=\langle 0 | \phi | 1 \rangle.\label{eq:fs}
\end{equation}

In contrast, for the potential energy we get contributions from $\Lambda_0=0$, which gives the disconnected part of the correlator, and $\Lambda_2$ which gives the asymptotic behaviour %of the connected correlator,
\begin{equation}
{G_V(t)}
= 
\langle  V({\phi})\rangle^2+
V_2^2e^{-\Lambda_2t} + \mathcal{O}(e^{-\Lambda_4t}),
\end{equation}
where
\begin{equation}
V_2=\langle 0 | V | 2 \rangle
=\int d\phi \psi_0(\phi)V({\phi}) \psi_2(\phi) 
.
\end{equation}

\subsection{Equal-time correlation function, power spectrum and spectral index}
\label{sec:23}
Because of de Sitter invariance, the equal-time correlation function between two different points in space can be obtained from the temporal correlation function $G_f(t)$, through
\be
G_f(0;\vec{x}_1,\vec{x}_2)=
G_f\left(\frac{2}{H}\ln (|\vec{x}_1-\vec{x}_2|H)\right),
\label{equ:timetospace}
\ee
which is valid at distances $|\vec{x}_1-\vec{x}_2|\gg 1/H$, and 
$\vec{x}=a\vec{r}$ is the physical, non-comoving coordinate. In the rest of the paper, we use only physical distances, because that makes the expressions for equal-time correlation functions time-independent.

Using Eq.~(\ref{equ:timetospace}), the spectral expansion (\ref{equ:spectral}) gives
\be
G_f(0;\vec{x}_1,\vec{x}_2)
=\sum_n \frac{f_n^2}{\left(|\vec{x}_1-\vec{x}_2|H
\right)^{2\Lambda_n/H}}.
\label{equ:spectralx}
\ee
At asymptotically long distances, the correlator therefore has a power-law form
\be
G_f(0;\vec{x})\sim \frac{A_f}{(|\vec{x}|H)^{n_f-1}},
\label{equ:powerlaw}
\ee
with constant parameters $A_f=f_n^2$ and 
\be n_f=1+\frac{2\Lambda_n}{H},\quad \text{for}\quad k\ll  H\,,\label{eq:n}
\ee for the lowest $n$ with $f_n\neq 0$.

In cosmology, equal-time correlation functions are often described in terms of their power spectrum ${\cal P}_f(k)$, defined by
\be
{\cal P}_f(k)=\frac{k^3}{2\pi^2}
\int d^3x e^{-i\vec{k}\cdot\vec{x}}G_f(0;\vec{x}).
\ee
Substituting Eq.~(\ref{equ:spectralx}) gives 
\be
{\cal P}_f(k)=\sum_n \f{2}{\pi}f^2_n\Gamma\bigg(2-2\f{\Lambda_n}{H}\bigg)\sin\bigg(\f{\Lambda_n\pi}{H}\bigg)\bigg(\f{k}{H}\bigg)^{2\Lambda_n/H}.
\label{eq:spectrumfull}
\ee

At long distances, $k\ll  H$, the power spectrum (\ref{eq:spectrumfull}) 
is also dominated by the leading term and has the power-law form,
\be
{\cal P}_f(k)\sim \frac{2}{\pi}A_f\Gamma\left[2-(n_f-1)\right]\sin\left(\frac{\pi (n_f-1)}{2}\right) \left(\frac{k}{H}\right)^{n_f-1}\approx
A_f \left(n_f-1\right) \left(\frac{k}{H}\right)^{n_f-1} ,\label{eq:spectrumP}
\ee
where the constants $A_f$ and $n_f$ are the same as in Eq.~(\ref{equ:powerlaw}),
and the last form is valid when $|n_f-1|\ll 1$.
In particular, this shows that $n_f$ is the spectral index, commonly defined
as
\be
\f{\ln{\cal P}_f(k)}{\ln k}= {n_f-1}.\label{eq:SI}
\ee

%%%%%%%%%%%%%%%%%%%%%%%%%%%%%%%%%%%%%%%%%%%%%%%%%%%%%%%%%%%%%%%%%%%%%%%%%%

\section{Example: a massive self-interacting field}
\label{sec:examples}

%%%%%%%%%%%%%%%%%%%%%%%%%%%%%%%%%%%%%%%%%%%%%%%%%%%%%%%%%%%%%%%%%%%%%%%%%%

\subsection{Eigenvalue equation}
\label{sec:eigenvalue}
%\label{sec:limiting_cases}

As an example of the formalism presented in the previous section we will discuss a potential with quadratic and quartic contributions
\begin{equation}
V(\phi)=\frac{1}{2}m^2\phi^2+\frac{\lambda}{4}\phi^4\,,\label{eq:pot2}
\end{equation}
with the assumption $m^2>0$. The analysis required for the double well potential, $m^2<0$, is significantly more complicated, which we will investigate in a separate publication \cite{M&R}. 

For the potential in Eq.~(\ref{eq:pot2}) the eigenvalue equation (\ref{e:sch}) becomes
\be
\f{1}{2}\bigg\{\f{\partial^2}{\partial\phi^2}
-\left(\frac{4\pi^2}{3H^4}\right)^2
\left(m^4\phi^2+2\lambda m^2\phi^4
+\lambda^2\phi^6
\right)
+\frac{4\pi^2}{3H^4}\left(m^2+3\lambda\phi^2\right)
%+\frac{2 \pi ^2 m^2}{3 H^4}-\frac{8 \pi ^4 m^4 \phi ^2}{9 H^8}+\frac{2 \pi ^2
%   \lambda  \phi ^2}{H^4}-\frac{16 \pi ^4 \lambda  m^2 \phi ^4}{9 H^8}&-\frac{8 \pi ^4 \lambda ^2 \phi ^6}{9 H^8}
\bigg\}\psi_n(\phi)%\nonumber%(m/H,\lambda;\phi)\nonumber
=-\f{4\pi^2}{H^3}\Lambda_n%\big(m/H,\lambda\big)
\psi_n(\phi)%(m/H,\lambda;\phi)
\,.
\ee
It is convenient to introduce a scaled version of the above equation expressed with only dimensionless parameters % from (\ref{eq:alpha})
\begin{align}
  &\bigg\{\frac{\partial^2}{\partial x^2}
  -U(\alpha;x)
  \biggr\}\psi_n(\alpha;x)
  =-8\pi^2\frac{\Lambda_n(\alpha)}{\lambda^{1/2}H}\psi_n(\alpha;x)
 % \frac{4 \pi ^2 \alpha}{3 \Omega ^2}+4 \pi ^2\frac{1-\frac{4\pi ^2}{9}\alpha^2 }{\Omega ^4}x^2-\frac{32 \pi ^4 \alpha x^4}{9 \Omega ^6}-\frac{16 \pi ^4 x^6}{9 \Omega ^8}+\frac{8 \pi ^2 {\left(1+\alpha\right)\tilde{\Lambda}_n\left(\alpha\right)}}{\Omega ^2}\bigg\}\psi_n(x)=0\
  ,\label{eq:quar}
\end{align}
where
\begin{equation}
x\equiv \f{\lambda^{1/4}}{H}\phi,\quad
\alpha\equiv\frac{m^2}{\sqrt{\lambda}H^2}
\,,\label{eq:alpha}
\end{equation}
and
\begin{equation}
    U(\alpha;x)
    =\left(\frac{4\pi^2}{3}\right)^2 x^2\left(\alpha+x^2\right)^2
  -\frac{4\pi^2}{3}\left(\alpha+3x^2\right).
\end{equation}
In this form it is apparent that up to an overall scale, the eigenvalues $\Lambda_n$ and the eigenfunctions $\psi_n$ depend only on one dimensionless parameter $\alpha$. In the next subsection, we will consider the limits of small and large $\alpha$ using perturbation theory, and the case of an arbitary $\alpha$ numerically.
From now on throughout this section we will drop the explicit $x$ dependence from the eigenfunctions. 

\subsection{Massless limit ($\alpha\ll 1$)}

Near the massless limit $m^2\ll \lambda^{1/2} H^2$, or equivalently $\alpha\ll 1$, we can find the eigenvalues and eigenfunctions as perturbative expansions in powers of $\alpha$,
\begin{equation}
    \Lambda_n(\alpha)=\Lambda_n^{(0)}+\alpha \Lambda_n^{(1)}+O(\alpha^2),
    \quad
    \psi_n(\alpha)=\psi_n^{(0)}+\alpha \psi_n^{(1)}+O(\alpha^2).
\end{equation}
To do this,
we Taylor expand Eq.~(\ref{eq:quar}) in powers of $\alpha$. This gives
\begin{equation}
    U(\alpha;x)
    =U^{(0)}(x)+\alpha U^{(1)}(x)+O(\alpha^2),
\end{equation}
where
\begin{equation}
    U^{(0)}(x)=U(0;x)=\left(\frac{4\pi^2}{3}\right)^2x^6-4\pi^2 x^2,
    \quad
    U^{(1)}(x)=2\left(\frac{4\pi^2}{3}\right)^2x^4-\frac{4\pi^2}{3}.
\end{equation}

The zeroth order case, $\alpha=0$, was first discussed in Ref.~\cite{Starobinsky:1994bd}. The lowest eigenvalues and eigenfunctions can be solved numerically with the overshoot/undershoot method from the eigenvalue equation
\be
  \bigg\{\frac{\partial^2}{\partial x^2}-U^{(0)}(x)\bigg\}\psi^{(0)}_n%=-\frac{8 \pi ^2 {\Lambda }_n}{\sqrt{\lambda}H}\psi_n(\phi)\equiv 
=
  \bigg\{\frac{\partial^2}{\partial x^2}+{4 \pi ^2  x^2}-\frac{16 \pi ^4 x^6}{9 }\bigg\}\psi^{(0)}_n%=-\frac{8 \pi ^2 {\Lambda }_n}{\sqrt{\lambda}H}\psi_n(\phi)\equiv 
  =-{8 \pi^2  } \frac{\Lambda^{(0)}_n}{\lambda^{1/2}H}\psi^{(0)}_n\,.%;\quad x\equiv\phi \f{\lambda^{1/4}}{H}\,.
  \label{eq:quar0}
\ee
The lowest five eigenvalues are
\begin{align}
\qquad\Lambda^{(0)}_0&=0\,,\nonumber\\
\Lambda^{(0)}_1%&\approx& 1.3685924536\sqrt{\frac{\lambda}{24\pi^2}}H
&\approx0.08892\sqrt{\lambda}H,\nonumber\\
\Lambda^{(0)}_2%&\approx& 4.4537088249\sqrt{\frac{\lambda}{24\pi^2}}H
&\approx0.28938\sqrt{\lambda}H,\nonumber\\
\Lambda^{(0)}_3%&\approx& 8.2596932608\sqrt{\frac{\lambda}{24\pi^2}}H,
&\approx0.53667\sqrt{\lambda}H\,,\nonumber\\
\Lambda^{(0)}_4%&\approx&12.7580695048\sqrt{\frac{\lambda}{24\pi^2}}H,\nonumber\\
%&\ldots&.
&\approx0.82895\sqrt{\lambda}H\,.
\label{eq:quartic}
\end{align}
By making use of a standard result from time independent perturbation theory for quantum mechanics the first order correction to the eigenvalue may be expressed as (see e.g. Ref.~\cite{griffiths2005introduction})
\begin{equation}
\Lambda_{n}^{(1)} \frac{8\pi^2}{\lambda^{1/2}H}=\langle \psi_n^{(0)}|U^{(1)}|\psi_n^{(0)}\rangle\,.\label{eq:QME}
%\int_{-\infty}^{\infty}\dd x\,\psi_n^{(0)}(x) U^{(1)}(x)\psi_{n}^{(0)}(x)
\end{equation}
With the above the $O(\alpha)$ corrections to the eigenvalues are given by
\begin{equation}
\Lambda_{n}^{(1)} 
%\frac{\lambda^{1/2}H}{8\pi^2}\langle \psi_n^{(0)}|U^{(1)}|\psi_n^{(0)}\rangle
%\int_{-\infty}^{\infty}\dd x\,\psi_n^{(0)}(x) U^{(1)}(x)\psi_{n}^{(0)}(x)
=
\frac{\lambda^{1/2}H}{6}\f{
\int_{-\infty}^{\infty}\dd x\,\psi_n^{(0)}\left[\frac{8\pi^2}{3}x^4 - 1\right]\psi_{n}^{(0)}}{
\int_{-\infty}^{\infty}\dd x\,\left(\psi_{n}^{(0)}\right)^2}\,.
\label{eq:smallAlphaPert}
\end{equation}
Eq. (\ref{eq:smallAlphaPert}) can only be evaluated analytically for the $n = 0$ eigenvalue, for which the result is zero (which is to be expected, since the perturbation does not change the normalisability of the zero eigenfunction, so the lowest eigenvalue must be zero). The results for the lowest five eigenvalues are given in Table \ref{tab:eigenvalues}.

\begin{table}%[htb]
	\centering
	\renewcommand{\arraystretch}{1.5}
	\begin{tabular}{|l|c|c|}
		\hline
		$\Lambda_n$ & $\alpha \ll 1$ & $\alpha \gg 1$ \\
		\hline
		$n=0$ & $0$ & $0$ \\
		\hline
		$n=1$ & $0.08892\sqrt{\lambda}H +  0.21478\sqrt{\lambda}H\alpha + O(\alpha^2)$ & \!$\frac{m^2}{H}\left(\frac{1}{3} + \frac{3}{8\pi^2\alpha^2}\right) + O\left(\frac{1}{\alpha^4}\right)$  \\
		\hline
		$n=2$ & $0.28938\sqrt{\lambda}H + 0.35152\sqrt{\lambda}H\alpha + O(\alpha^2)$ & \!$\frac{m^2}{H}\left(\frac{2}{3} + \frac{3}{2\pi^2\alpha^2}\right) + O\left(\frac{1}{\alpha^4}\right)$  \\
		\hline
		$n=3$ & $0.53667\sqrt{\lambda}H + 0.52930\sqrt{\lambda}H\alpha + O(\alpha^2)$ & $\frac{m^2}{H}\left(1+ \frac{27}{8\pi^2\alpha^2}\right) + O\left(\frac{1}{\alpha^4}\right)$  \\
		\hline
		$n=4$ & $0.82895\sqrt{\lambda}H + 0.70696\sqrt{\lambda}H\alpha + O(\alpha^2)$& $\frac{m^2}{H}\left(\frac{4}{3} + \frac{6}{\pi^2\alpha^2}\right) + O\left(\frac{1}{\alpha^4}\right)~$  \\
		\hline
	\end{tabular}
	\caption{\label{tab:eigenvalues}Lowest eigenvalues $\Lambda_n$ %for the $\frac{1}{2}m^2\phi^2 + \frac{1}{4}\lambda\phi^4$ potential 
	in the limits of large and small $\alpha = {m^2}/{\lambda^{1/2}H^2}$.}
\end{table}

One can also calculate the coefficients (\ref{eq:fn}) for $f(\phi)=\phi^j$ to zeroth order in $\alpha$ from the integral
\be
(\phi^j)_n=
\int d\phi\, \psi_0^{(0)}(\phi)\phi^j\psi_n^{(0)}(\phi)+O(\alpha),
\label{eq:coeffzeroth}
\ee
using the numerically calculated zeroth order eigenfunctions $\psi_n^{(0)}$
. For the lowest $j$ and $n$, values are
given in Table~\ref{tab:coeffslambda}.

\begin{table}[]
    \centering
    \renewcommand{\arraystretch}{1.5}
    \begin{tabular}{|c|c|c|c|c|}
\hline
$|(\phi^j)_n|$ & $j=1$ & $j=2$ & $j=3$ & $j=4$ \\
        \hline
         $n=1$ & $0.36016\,\lambda^{-1/4}H$ & $0$ & $0.09608(\lambda^{-1/4}H)^3$ & $0$\\
         \hline
         $n=2$ & $0$ & $0.13929(\lambda^{-1/4}H)^2$ & $0$ & $0.06046(\lambda^{-1/4}H)^4$\\
         \hline
         $n=3$ & $0.04475\,\lambda^{-1/4}H$ & $0$ & $0.07203(\lambda^{-1/4}H)^3$ & $0$\\
         \hline
         $n=4$ & $0$ & $0.03435(\lambda^{-1/4}H)^2$ & $0$ & $0.04271(\lambda^{-1/4}H)^4$
       \\ \hline
    \end{tabular}
    \caption{Numerically computed coefficients $|(\phi^j)_n|$ in the massless self-interacting limit ($\alpha=0$).}
    \label{tab:coeffslambda}
\end{table}

\subsection{Free limit ($\alpha\gg 1$)}
In the opposite limit $\lambda\ll (m/H)^4$, or equivalently $\alpha\gg 1$, the 
coefficients of Eq.~(\ref{eq:quar}) diverge, so it is convenient to rescale again,
and define 
\begin{equation}
    \tilde{x}=\alpha^{1/2}x=\frac{m}{H^2}\phi.
\end{equation}The eigenvalue equation then becomes
\begin{equation}
\bigg\{\frac{\partial^2}{\partial \tilde{x}^2}
  -\tilde{U}(\alpha;\tilde{x})
  \biggr\}\psi_n(\alpha)
  =-8\pi^2\frac{\Lambda_n(\alpha)}{\alpha\lambda^{1/2}H}\psi_n(\alpha),
\end{equation}
where
\begin{equation}
    \tilde{U}(\alpha;\tilde{x})
    =\left(\frac{4\pi^2}{3}\right)^2 \tilde{x}^2\left(1+\frac{\tilde{x}^2}{\alpha^2}\right)^2
  -\frac{4\pi^2}{3}\left(1+\frac{3\tilde{x}^2}{\alpha^2}\right).
\end{equation}

We want to solve this perturbatively in powers of $\alpha^{-2}$,
so we write
\begin{equation}
    \Lambda_n(\alpha)=\Lambda^{(0)}_n+\frac{1}{\alpha^2}\Lambda^{(1)}_n+O\left(\frac{1}{\alpha^4}\right),\quad
    \psi_n(\alpha)=\psi^{(0)}_n+\frac{1}{\alpha^2}\psi^{(1)}_n+O\left(\frac{1}{\alpha^4}\right),
\end{equation}
and expand the potential as
\begin{equation}
\tilde{U}(\alpha;\tilde{x})=
\tilde{U}^{(0)}(\tilde{x})+
\frac{1}{\alpha^2}\tilde{U}^{(1)}(\tilde{x})+
O\left(\frac{1}{\alpha^4}\right),
\end{equation}
where
\begin{equation}
    \tilde{U}^{(0)}(\tilde{x})
    =\left(\frac{4\pi^2}{3}\right)^2 \tilde{x}^2-\frac{4\pi^2}{3}
    ,\quad
    \tilde{U}^{(1)}(\tilde{x})
    =2\left(\frac{4\pi^2}{3}\right)^2 \tilde{x}^4-4\pi^2\tilde{x}^2.
\end{equation}
The zeroth order equation is nothing more than the standard equation for a harmonic oscillator, and
eigenfunctions can be written in terms of the Hermite polynomials as\footnote{$H_n(x) = (-1)^n e^{x^2}\frac{d^n}{dx^n}e^{-x^2}$}
%{\color{green}\bf Write in terms of $\tilde{x}$?}
\begin{equation}
\psi^{(0)}_n=\frac{\sqrt{m}}{H}\frac{1}{\sqrt{2^n n!}}\bigg(\frac{4 \pi }{3 }\bigg)^{1/4} e^{-\frac{2 \pi ^2 \tilde{x}^2}{3}}H_n\left(\frac{2 \pi  \tilde{x}}{\sqrt{3}}\right)
%\psi^{(0)}_n(\phi)=\frac{1}{\sqrt{2^n n!}}\bigg(\frac{4 \pi  m^2}{3 H^4}\bigg)^{1/4} e^{-\frac{2 \pi ^2 m^2 \phi^2}{3 H^4}}H_n\left(\frac{2 \pi  m \phi}{\sqrt{3} H^2}\right)
\,,\label{eq:fulls}
\end{equation}
corresponding to the zeroth-order eigenvalues
\begin{equation}
\Lambda^{(0)}_n = %\tilde{\Lambda}_n(\infty)\frac{m^2}{H}=
\f{n}{3}\alpha\lambda^{1/2}H=
\f{n}{3}\frac{m^2}{H}\,.\label{eq:quadratic}
\end{equation}
The perturbative correction to the eigenvalues again comes via Eq.~(\ref{eq:QME})
\begin{align}
\Lambda_n^{(1)} %&=& \frac{\alpha\lambda^{1/2}H}{8\pi^2}\langle \psi_n^{(0)} | \tilde{U}^{(1)} | \psi_n^{(0)} \rangle
%\int_{-\infty}^{\infty}\dd \tilde{x}\,\psi_n^{(0)}(\tilde{x})\tilde U^{(1)}(\tilde{x})\psi_n^{(0)}(\tilde{x})
%\nonumber\\
%&=&
=
\frac{m^2}{8\pi^2H} \frac{1}{\sqrt{\pi}2^n n!}
\int_{-\infty}^{\infty}\dd u\, H_n(u)^2(2u^4-3u^2)e^{-u^2}=\frac{3  n^2}{8 \pi ^2 }\frac{m^2}{H}.
\label{eq:harmonic0}
\end{align}
The first five eigenvalues and their corrections are given in Table \ref{tab:eigenvalues}.

Again, one can calculate the coefficients (\ref{eq:fn}) to zeroth order in $\alpha^{-2}$ using the integral~(\ref{eq:coeffzeroth}). This gives
\be
(\phi^j)_1=
\begin{dcases}
\sqrt{\frac{2}{\pi}}\left(\frac{3H^4}{4\pi^2m^2}\right)^{j/2}
\Gamma\left(\frac{j}{2}+1\right)+O(\alpha^{-2}), &\text{for odd $j$},\\
0+O(\alpha^{-2}),&\text{for even $j$},
\end{dcases}
%\int d\phi\, \psi_0^{(0)}(\phi)\phi^j\psi_1^{(0)}(\phi)
%=
%\left(\frac{H^2}{m}\right)^{2j+1}
%=\sqrt{\frac{3}{2\pi^3}}\left(\frac{3}{4\pi^2}\right)^j\Gamma\left[\frac{3}{2}+j\right]
%\left(\frac{H^2}{m}\right)^{2j+1}
%
\label{equ:coeffsodd}
\ee
and
\be
(\phi^{j})_2=
%\int d\phi\, \psi_0^{(0)}(\phi)\phi^{j}\psi_2^{(0)}(\phi)
\begin{dcases}
0+O(\alpha^{-2}), & \text{for odd $j$},\\
\frac{j}{2}\sqrt{\frac{2}{\pi}}\left(\frac{3H^4}{4\pi^2m^2}\right)^{j/2}\Gamma\left(\frac{j+1}{2}\right)+O(\alpha^{-2}),
& \text{for even $j$.}
\end{dcases}
%\left(\frac{H^2}{m}\right)^{2j}
%.
\label{equ:coeffseven}
\ee

In principle it is possible to calculate perturbative corrections to the eigenfunctions. However, we will omit this from our discussion as they are of limited use due to their analytically involved nature and limited applicability \cite{griffiths2005introduction}, especially since as we demonstrate in the next section full numerical solutions are readily available.

%%%%%%%%%%%%%%%%%%%%%%%%%%%%%%%%%%%%%%%%%%%%%%%%%%%%%%%%%%%%%%%%%%%%%%%%%%

\subsection{Arbitrary $\alpha$}
\label{numerics}
\begin{figure}\begin{center}
\includegraphics[width=0.9\textwidth,trim={0cm 0 0 2cm},clip]{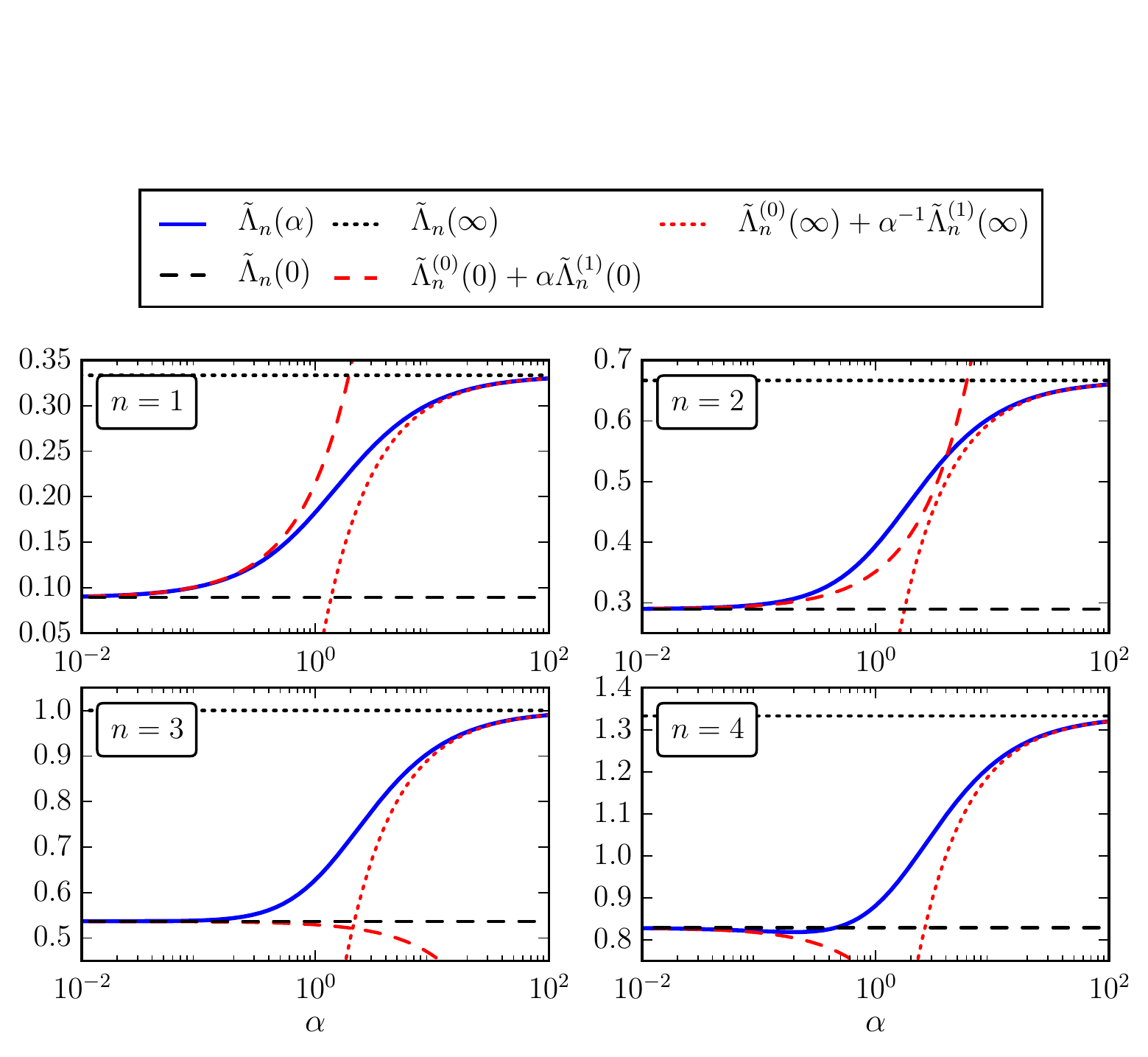}
\end{center}
\caption{\label{fig:eigenvalues}
The dimensionless eigenvalues from {Eq.}~(\ref{scaledlambda}) for the potential (\ref{eq:pot2}) solved from {Eq.}~(\ref{eq:quar}) as a function of $\alpha\equiv m^2/(H^2\sqrt{\lambda})$. The interpolation between quadratic (\ref{eq:quadratic}) and quartic (\ref{eq:quartic}) results is clearly evident. The red curves correspond to the leading perturbative corrections, which can be obtained from Table \ref{tab:eigenvalues} with Eq.~(\ref{scaledlambda}).}
\end{figure}

For a generic value of $\alpha$, away from the two limits, 
the eigenvalue equation (\ref{eq:quar}) can be solved numerically.
It is convenient to do a further rescaling,
%{\color{green}\bf Note I have introduced $z$ for the numerical field value}
\begin{equation}
    z\equiv \Omega x\equiv \f{\lambda^{1/4}\Omega}{H}\phi, \quad \Omega\equiv\bigg(1+\frac{m}{H\lambda^{1/4}}\bigg)\equiv\left(1+\sqrt{\alpha}\right),\label{eq:fsca}
\end{equation}
so that the equation becomes
\begin{align}
  &\bigg\{\frac{\partial^2}{\partial z^2}+\frac{4 \pi ^2 \alpha}{3 \Omega ^2}+4 \pi ^2\frac{1-\frac{4\pi ^2}{9}\alpha^2 }{\Omega ^4}z^2-\frac{32 \pi ^4 \alpha z^4}{9 \Omega ^6}-\frac{16 \pi ^4 z^6}{9 \Omega ^8}+\frac{8 \pi ^2 {\left(1+\alpha\right)\tilde{\Lambda}_n\left(\alpha\right)}}{\Omega ^2}\bigg\}\psi_n(z)=0\,,\label{eq:quarz}
\end{align}
where
\begin{equation}
    \tilde{\Lambda}_n(\alpha)=\frac{\Lambda_n}{\lambda^{1/2} H (1+\alpha)}=\frac{\Lambda_n}{\lambda^{1/2}H + m^2/H}.
    \label{scaledlambda}
\end{equation}
\begin{figure}\begin{center}
\includegraphics[width=0.9\textwidth]{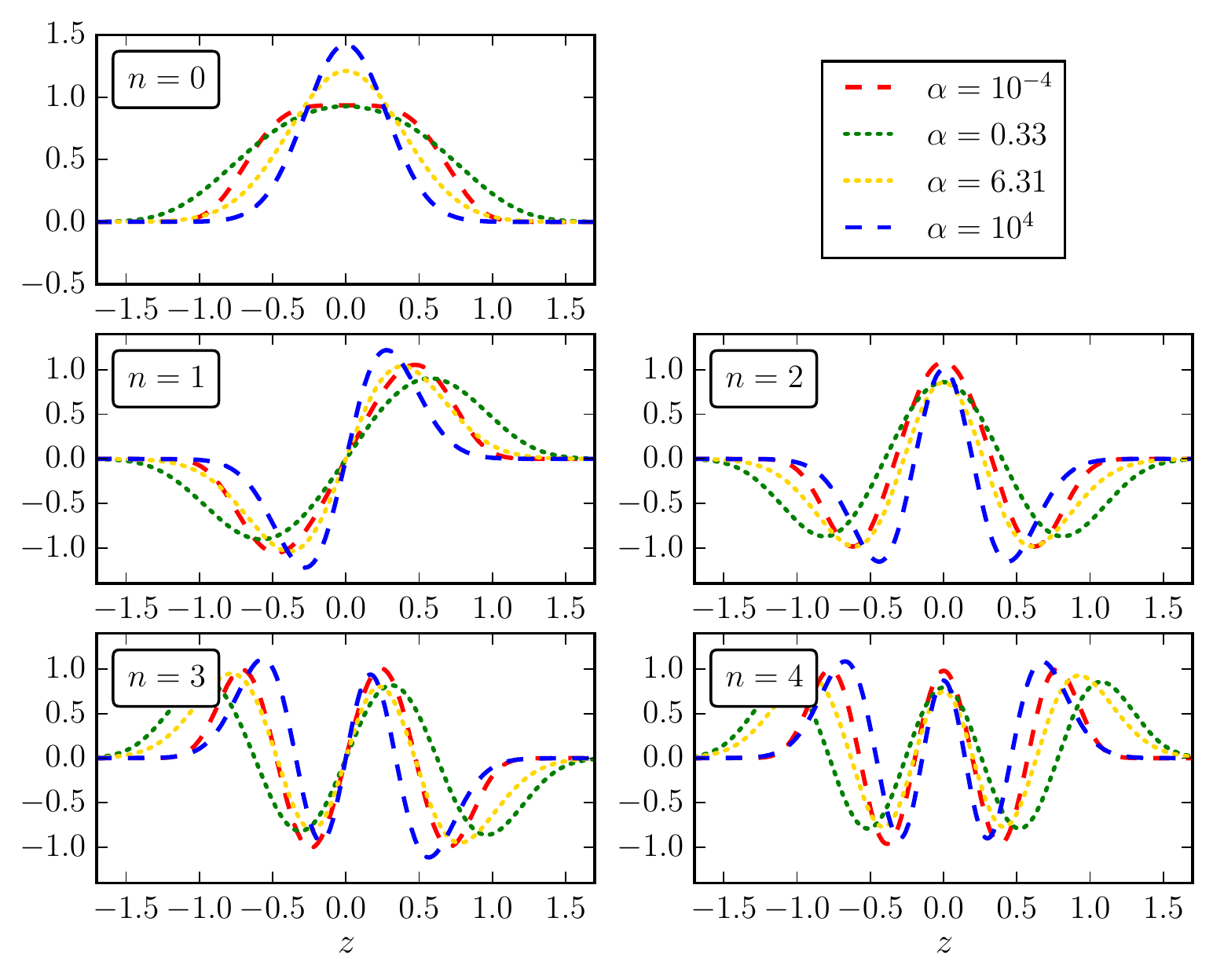}
\end{center}
\caption{\label{fig:eigenfunctions}
The scaled eigenfunctions $\tilde{\psi}_n(\alpha;z)$ defined in {Eq.}~(\ref{eq:scaledefs}) for the potential (\ref{eq:pot2}) solved from (\ref{eq:quarz}) by making use of the definitions (\ref{eq:fsca}) and (\ref{scaledlambda}). The quartic and quadratic limits are reached at $\alpha=0$ and $\alpha=\infty$, respectively.}
\end{figure}

The eigenvalues and -functions from {Eq.}~(\ref{eq:quarz}) can be solved for example with the overshoot/undershoot method.\footnote{When calculating the eigenfunctions with the overshoot/undershoot method we have truncated the solutions to a finite range $z\in[-z_0,z_0]$ surrounded by $\psi_n(z)=0$ where the cut-off $z_0$ is set by finding when the numerical solution $\psi_n(z)$ is closest to zero before diverging.} The results for the first four non-zero eigenvalues are presented in Fig. \ref{fig:eigenvalues}, which clearly coincide with {Eqs.}~(\ref{eq:quadratic}) and (\ref{eq:quartic}) as the limiting cases. In a similar fashion one may introduce scaled eigenfunctions satisfying
\begin{equation}
    \psi_n\equiv \sqrt{\frac{\lambda^{1/4}\Omega}{H}}\tilde{\psi}_n(\alpha)\quad\Rightarrow\quad \int^{\infty}_{-\infty} d\phi\, | \psi_n|^2=\int^{\infty}_{-\infty} dz\, | \tilde{\psi}_n(\alpha)|^2=1\,, \label{eq:scaledefs}
\end{equation}
where we have again dropped the the explicit $z$ and $\phi$ dependences from the eigenfunctions.
%with $\Omega$ and $x$ from (\ref{eq:eigenScaling0}-\ref{eq:eigenScaling}).
The results for the first five eigenfunctions are presented in Fig. \ref{fig:eigenfunctions}.

\begin{figure}\begin{center}
		\includegraphics[width=1.0\textwidth,trim={0cm 0 0 5.9cm},clip]{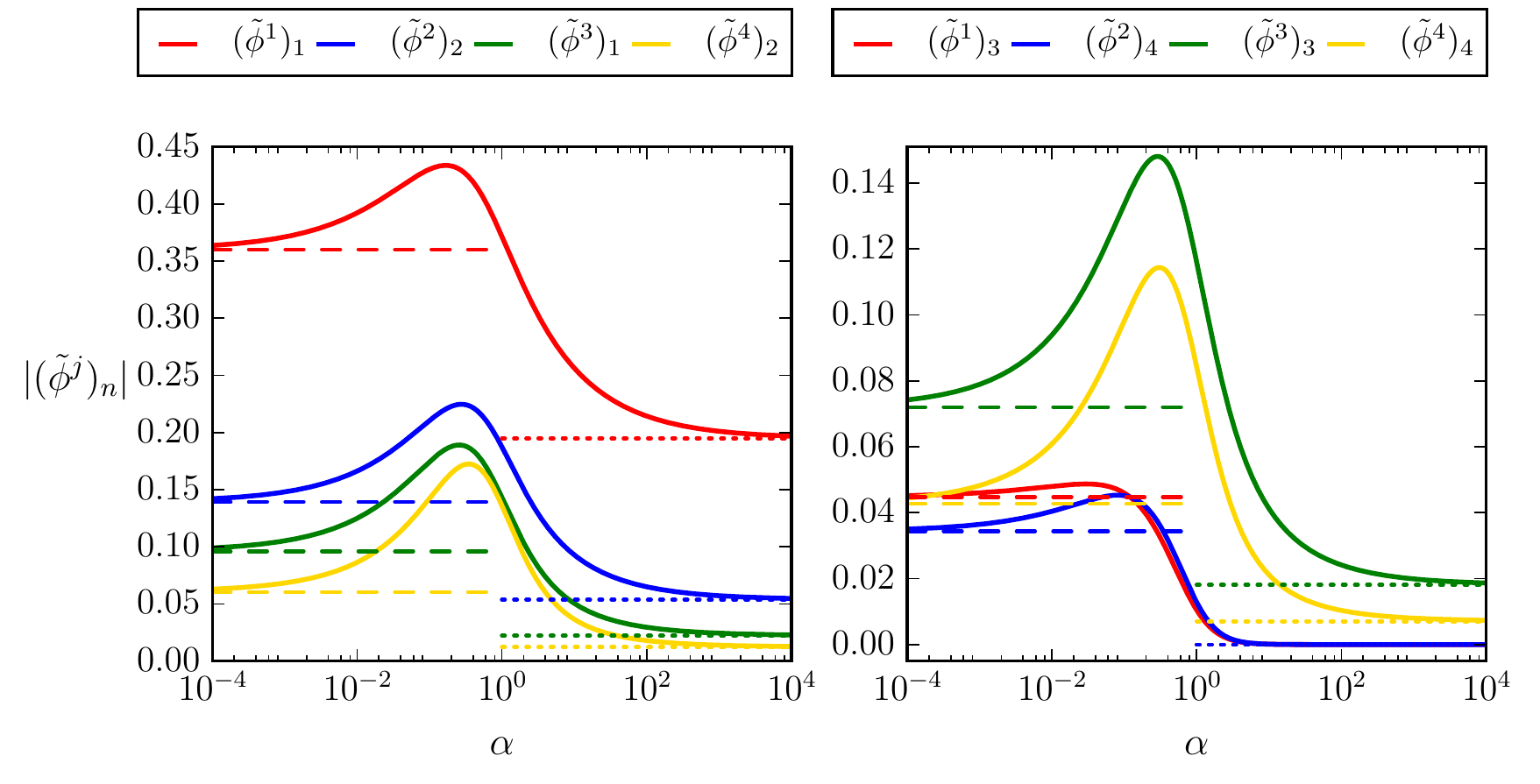}
	\end{center}
	\caption{\label{fig:fs}
		The leading terms in the spectral expansion when $f(\phi)=\phi^j$ with $j\leq4$, scaled according to {Eq.}~(\ref{eq:fsa}). The dashed and dotted lines correspond to the quartic $(\alpha\rightarrow0)$ and quadratic $(\alpha\rightarrow\infty)$  limits, respectively, given in~Table~\ref{tab:coeffslambda} and Eqs.~(\ref{equ:coeffsodd}) and (\ref{equ:coeffseven})  for the leading results (left).}
\end{figure}

In order to calculate correlators with the spectral expansion (\ref{equ:spectral}) one needs expressions for the coefficients $f_n$ defined in {Eq.}~(\ref{eq:fn}). They depend on the specific form of $f(\phi)$ and thus cannot be given independently of the correlator unlike the eigenvalues. As an important special case sufficient for most applications here we show the leading $f_n$ contributions when $f(\phi)$ is a monomial $\phi^j$, with $j\leq 4$. A convenient dimensionless version of the polynomial coefficients comes with the help of {Eq.}~(\ref{eq:fsca})
\begin{equation}
(\tilde{\phi}^j)_n\equiv\bigg(\f{\lambda^{1/4}\Omega}{H}\bigg)^j({\phi}^j)_n=\bigg(\f{\lambda^{1/4}\Omega}{H}\bigg)^j\int^{\infty}_{-\infty} d\phi \psi_0\phi^j  \psi_n
=\int^{\infty}_{-\infty} d z\, \tilde{\psi_0}(\alpha)z^j  \tilde{\psi}_n(\alpha)\,,\label{eq:fsa}
\end{equation}
which are shown in Fig. \ref{fig:fs}. The quartic and quadratic limits for the first non-zero coefficients are given in Table~\ref{tab:coeffslambda} and Eqs.~(\ref{equ:coeffsodd}) and (\ref{equ:coeffseven}).
%The next-to-leading corrections are
%\be
%|(\tilde{\phi}^j)_n|\overset{\alpha\rightarrow0}{=}\begin{dcases}&0.04475\,;\quad j=1,n=3 \\
%&0.03435\,;\quad j=2,n=4 \\
%&0.07205\,;\quad j=3,n=3\\
%&0.04271\,;\quad j=4,n=4 \\\end{dcases}\,;\quad |(\tilde{\phi}^j)_n|\overset{\alpha\rightarrow\infty}{=}\begin{dcases}&0\,;\quad\phantom{.01814} j=1,n=3 \\
%&0\,;\quad\phantom{.01814} j=2,n=4 \\
%&0.01814\,;\quad j=3,n=3 \\
%&0.00707\,;\quad j=4,n=4 \\\end{dcases}\,.\label{eq:limitP}
%\ee

%%%%%%%%%%%%%%%%%%%%%%%%%%%%%%%%%%%%%%%%%%%%%%%%%%%%%%%%%%%%%%%%%%%%%%%%%%
\section{Validity of common approximations}
\label{sec:comparison}
\subsection{Full results}
As an illustration of the presented formalism and how it relates to other often implemented approximations, we
consider equal-time correlators of the field $\phi$ and its density contrast
\be{\delta}\equiv \f{\delta\rho_\phi}{\langle{\rho}_\phi\rangle}=\f{V({\phi})-\langle V({\phi})\rangle}{\langle\label{eq:DC} V({\phi})\rangle}\,,\ee
where as usual in cosmology the fluctuation is defined as the difference to the mean $\delta f({\phi}) \equiv  f({\phi})-\langle f({\phi})\rangle$ and we have neglected the kinetic term in the energy density.

Assuming that the energy density $\rho_\phi$ is an even function of the field $\phi$,
Eq.~(\ref{equ:spectralx}) shows that at long distances, the correlators are given by
\begin{eqnarray}
\langle{\phi}(0){\phi}(\vec{x})\rangle&\overset{|\vec{x}| H \gg1}{\approx}&
\f{(\phi_1)^2}{(|\vec{x}| H)^{2\Lambda_1/H}}\,,
\label{eq:contphi}
\\
\langle{\delta}(0){\delta}(\vec{x})\rangle&\overset{|\vec{x}| H \gg1}{\approx}&\label{eq:contF}
\f{(\delta_2)^2}{(|\vec{x}| H)^{2\Lambda_2/H}}\,,
\end{eqnarray}
where $\phi_1$ and $\delta_2$ are given by Eq.~(\ref{eq:fn}), and $\Lambda_1$ and $\Lambda_2$ are eigenvalues obtained from Eq.~(\ref{e:sch}).
In the full stochastic approach, the spectral indices $n_f$ and amplitudes $A_f$ defined by Eq.~(\ref{equ:powerlaw}) are therefore
\begin{equation}
n_\phi=1+2\frac{\Lambda_1}{H}
,\quad A_\phi=(\phi_1)^2\,,
\label{eq:nAgeneral0}
\end{equation}
for the field and
\begin{equation}
n_\delta=1+2\frac{\Lambda_2}{H},\quad 
A_\delta=(\delta_2)^2\,,
\label{eq:nAgeneral}
\end{equation}
for the density contrast. The corresponding power spectra are given by Eq.~(\ref{eq:spectrumP}) as
\begin{eqnarray}
{\cal P}_\phi(k)&\overset{ k/H \ll 1}{=}&
%\frac{2}{\pi}(\phi_1)^2\Gamma\left(2-2\frac{\Lambda_1}{H}\right)\sin\left(\frac{\Lambda_1\pi}{H}\right)\left(\frac{k}{H}\right)^{2\Lambda_1/H}
\approx\frac{2\Lambda_1}{H}(\phi_1)^2\left(\frac{k}{H}\right)^{2\Lambda_1/H}
,
\nonumber\\
{\cal P}_\delta(k)&\overset{ k/H \ll 1}{=}&
%\frac{2}{\pi}(\delta_2)^2\Gamma\left(2-2\frac{\Lambda_2}{H}\right)\sin\left(\frac{\Lambda_2\pi}{H}\right)\left(\frac{k}{H}\right)^{2\Lambda_2/H}
\approx\frac{2\Lambda_2}{H}(\delta_2)^2\left(\frac{k}{H}\right)^{2\Lambda_2/H},
\label{equ:Presults}
\end{eqnarray}
where we have assumed $\Lambda_{1,2}\ll H$.

For a quadratic potential, $V(\phi)=(m^2/2)\phi^2$,
the spectral coefficients are given by Eqs.~(\ref{equ:coeffsodd})
 and (\ref{equ:coeffseven}) and the eigenvalues by {Eq.}~(\ref{eq:quadratic}). They give
\begin{eqnarray}
n_\phi= 1+\frac{2}{3}\frac{m^2}{H^2},\quad A_\phi&=&
(\phi_1)^2=\frac{3}{8\pi^2}\frac{H^4}{m^2},
\label{equ:quadfield}
\\n_\delta =  1+\frac{4}{3}\frac{m^2}{H^2},\quad 
A_\delta&=&
(\delta_2)^2 = \left(\f{(\phi^2)_2}{\langle \phi^2\rangle}\right)^2=
 2\,.\label{eq:nquadratic0}
\end{eqnarray}
%The spectral indices are (see Eq.~(\ref{eq:quadratic})) %(\ref{eq:spectrumP})--(\ref{eq:n}))
%\be
%n_\phi= 1+\frac{2}{3}\frac{m^2}{H^2},\quad
%n_\delta =  1+\frac{4}{3}\frac{m^2}{H^2}\,,
%\label{eq:nquadratic}
%\ee
The full expressions for the power spectra are
\be
{\cal P}_\phi=\frac{H^2}{4\pi^2}\left(\frac{k}{H}\right)^{2m^2/3H^2}
,\quad
{\cal P}_\delta={\frac{8m^2}{3H^2}}\left(\frac{k}{H}\right)^{4m^2/3H^2}.
\label{equ:Pfree}
\ee

For a quartic potential, $V(\phi)=(\lambda/4)\phi^4$, 
the spectral coefficients are given in Table~\ref{tab:coeffslambda} and the eigenvalues by {Eq.}~(\ref{eq:quartic}) which lead to
% and (\ref{eq:potV}),
\begin{eqnarray}
n_\phi\approx 1+0.17784\lambda^{1/2},\quad A_\phi&=&(\phi_1)^2=0.12972\frac{H^2}{\lambda^{1/2}},
\label{equ:quartfield}
\\
n_\delta \approx  1+0.57876\lambda^{1/2},\quad A_\delta&=&(\delta_2)^2 = \left(\f{(\phi^4)_2}{\langle \phi^4\rangle}\right)^2=
\left(\f{8\pi^2}{3}(\tilde{\phi}^4)_2\right)^2\approx2.53198\,.
\label{eq:nquartic}
\end{eqnarray}
%The spectral indices are (see Eq.~(\ref{eq:quartic})) %(\ref{eq:spectrumP})--(\ref{eq:n}))
%\be
%n_\phi\approx 1+0.17784\lambda^{1/2},\quad
%n_\delta \approx  1+0.57876\lambda^{1/2}\,,
%\label{eq:nquartic}
%\ee
%with the numerical value from ().
The full expressions for the power spectra are then
%{\color{green} (AR: Do we have $\phi_3$ and $\delta_4$? Then we could give the subleading terms, too.)}
\begin{align}
{\cal P}_\phi&\approx 0.0231\,H^2\left(\frac{k}{H}\right)^{0.17784\lambda^{1/2}}+0.0021\,H^2\left(\frac{k}{H}\right)^{1.07334\lambda^{1/2}}
+\ldots
, \nonumber \\
{\cal P}_\delta&\approx 1.4655\,\lambda^{1/2}\left(\frac{k}{H}\right)^{0.57876\lambda^{1/2}}+2.0949\,\lambda^{1/2}\left(\frac{k}{H}\right)^{1.65790\lambda^{1/2}}+\ldots,
\label{equ:Pmassless}
\end{align}
where for completeness we have also included the next-to-leading order terms. From them we can see that the leading term dominates the field spectrum ${\cal P}_\phi$ on all superhorizon scales, but in the density spectrum ${\cal P}_\delta$ only when $k\lesssim 10^{-0.144\,\lambda^{-1/2}} H.$ If $\lambda$ is sufficiently small, this can become important.

The results for the asymptotic amplitude $A_\delta$ and the spectral index $n_\delta$ for the density contrast when the potential has a quartic and a quadratic term (\ref{eq:pot2}) are shown in Fig.~\ref{fig:delta}.

\subsection{Approximation schemes}
{
It is instructive to compare the stochastic results 
with different approximations used in the literature.
In the following we will consider four such approximations, which can be thought of as the four different outcomes of two separate binary choices:  
{The first choice is how to express the density contrast correlator in terms of the field correlators.
For this we consider two approaches, which were refer to as the {\em mean field (MF)} and {\em Gaussian} approximations.
The second choice is how to compute the field correlator, and for that we consider the conventional {\em linear} approximation and the {\em stochastic} approach. We will refer to the resulting {four} approximations as the linear-MF,
stochastic-MF ,
linear-Gaussian, and 
stochastic-Gaussian 
approximations.
In Fig.~\ref{fig:delta} we
compare the amplitude and the spectral index of the density contrast computed using the full stochastic approach (\ref{eq:nAgeneral}) to these four different approximations. We find that all of the approximations consistently overestimate the amplitude and underestimate the spectral tilt, leading to an overall overestimation of power on cosmological scales.}

\subsubsection{Linear-MF approximation}
One way some authors have have tried to {compute the density contrast correlator is by treating the}
%interacting case is to
%treat the 
root-mean-square value of the field, $\varphi\equiv\left(\langle{\phi}^2\rangle\right)^{1/2}$, as if it was 
{constant homogeneous background field, and expanding the field fluctuations around it}
%the mean value, by writing
\be
\phi=\varphi+\delta\phi,\quad \delta\phi\ll \varphi.
\ee
{We refer to this as the mean field (MF) approximation.}
The density contrast is then {expressed} %expanded naively 
to linear order in $\delta\phi$ as
\be
\delta^{\rm MF}=\frac{V'(\varphi)}{V(\varphi)}\delta\phi,
\label{eq:lin2def}
\ee
and therefore its correlator would be
\be
\langle{\delta}^{\rm MF}(0){\delta}^{\rm MF}(\vec{x})\rangle
=\left(\frac{V'(\varphi)}{V(\varphi)}\right)^2\langle{\delta\phi}(0){\delta\phi}(\vec{x})\rangle.
\label{eq:linrelation}
\ee
This would imply
\be
n_\delta^{\rm MF}=n_\phi,
\quad
A_\delta^{\rm MF}=\left(\frac{V'(\varphi)}{V(\varphi)}\right)^2 A_\phi
=4
\frac{\left(1+\frac{\lambda}{m^2}\varphi^2\right)^2}{\left(1+\frac{\lambda}{2m^2}\varphi^2\right)^2}\frac{A_\phi}{\varphi^2}
.
\label{equ:MFrel2}
\ee

{The next step is to compute the field correlator and, hence, the parameters $n_\phi$, $A_\phi$ and $\varphi$.
This requires a second choice of an approximation.}
In the linear-MF approximation the field correlator is taken to be the linear solution to the equation of motion, which then means that the different comoving modes decouple.
%simplest form of the linear approximation neglects interactions all together. As it becomes exact in the free limit it is a useful check to compare the stochastic results with that case.} 
By solving the linear mode functions exactly one obtains the 2-point correlator for the field~\cite{Chernikov:1968zm,Bunch:1978yq,Birrell:1982ix}
\begin{align}
\langle{\phi}(0){\phi}(\vec{x})\rangle^{\rm lin}&=\f{H^2(1-c)(2-c)}{16\pi \sin\left(\pi(1-c)\right)}\,_2F_1\bigg[3-c,c;2;1-\bigg(\f{\vec{x}H}{2}\bigg)^2\bigg]\,; \qquad c=\f32 -\sqrt{\f94 -\f{M^2{(\varphi)}}{H^2}}\nonumber \\ &\overset{|\vec{x}|H\rightarrow\infty}{=}\f{H^2(1-c)(2-c)4^c}{16\pi \sin\left(\pi(1-c)\right)} \frac{\Gamma (3-2 c)}{\Gamma (2-c) \Gamma (3-c)}|\vec{x}H|^{-2c}
%\nonumber\\
%&=\frac{3H^4}{8\pi^2m^2}\left(1
%+\frac{(3\ln 4-7)m^2}{9H^2}+O\left(\frac{m^4}{H^4}\right)
%\right)
%\left(|\vec{x}|H\right)^{-2c}%\bigg(1+{\cal O}\bigg(\frac{m^2}{H^2}\bigg)\bigg)
\,,
\label{eq:linearisedsol}
\end{align}
with the effective mass
\be
M^2{(\varphi)}=V''(\varphi)=m^2+3\lambda\varphi^2\,.
\ee
The %{\color{red} background field value $\varphi$}
field variance 
$\langle{\phi}^2\rangle$ 
can be obtained from Eq.~(\ref{eq:linearisedsol}). The full expression is ultraviolet divergent, but the leading term in powers of $M^2/H^2$ is finite, %which can be obtained e.g. via point splitting $|\vec{x} H|\sim\epsilon$
%The long wavelength contribution to the variance comes via the standard Bunch-Davies modes by introducing a cut-off at the horizon
\be \label{eq:var}
{\varphi^2}=
\langle{\phi}^2\rangle^{\rm lin}
=\frac{(2-c) (c-1) H^2 \left(H_{2-c}+\psi ^{(0)}(c)+\gamma_e -1\right)}{16 \pi ^2}=\frac{3}{8\pi^2}\frac{H^4}{M^2{(\varphi)}}\left(1
+{\cal O}\left(\frac{M^2}{H^2}\right)
%-\frac{7}{9}\frac{M^2}{H^2}
%+{\cal O}\left(\frac{M^4}{H^4}\right)
\right)\,,
%=H^2\int_{0}^{1} \frac{d x\,x^2 }{8\pi} |H^{(1)}_{3/2-c}(x)|^2  \approx {H}^2\frac{\Gamma[3/2-c]^2}{2^{1+2c}\pi^3 c}=\frac{3}{8\pi^2}\frac{H^4}{m^2}\bigg(1+{\cal O}\bigg(\frac{m^2}{H^2}\bigg)\bigg)\,.
\ee
%where in the second step we have used the asymptotic form of the Hankel function. 
where $H_n$ is a harmonic number and $\psi^{(i)}(z)$ the polygamma function. 
{The background field value $\varphi$ is obtained by solving this equation.}

The spectral index and amplitude for the field correlator can be read off from Eq.~(\ref{eq:linearisedsol})
\begin{align}
n_\phi^{\rm lin}-1&=2c%4-\sqrt{9-4\frac{M^2}{H^2}}
=\frac{2}{3}\frac{M^2{(\varphi)}}{H^2}
\left(1+\frac{1}{9}\frac{M^2{(\varphi)}}{H^2}+\cdots%{\cal O}\left(\frac{M^4}{H^4}\right)
\right),\nonumber \\ A_\phi^{\rm lin}&=\frac{3}{8\pi^2}\frac{H^4}{M^2{(\varphi)}}\left(1
+\frac{(3\ln 4-7)}{9}\frac{M^2{(\varphi)}}{H^2}+\cdots%{\cal O}\left(\frac{M^4}{H^4}\right)
\right)%,\nonumber\\
.\label{eq:nphilin}
\end{align}
Comparison with Eqs.~(\ref{eq:nquadratic0}) shows that in the quadratic limit ($\alpha\rightarrow\infty$ or equivalently $\lambda\rightarrow0$) the stochastic results agree with the above only to linear order in $m^2/H^2$. This is because its starting point, Eq.~(\ref{eq:Langevin}), relies on the slow roll approximation \cite{Starobinsky:1994bd}. %(\ref{eq:linearisedsol}) with the replacement $m^2\rightarrow V''(\varphi)=m^2+3\lambda\varphi^2$, so that

The spectral index and amplitude for the density contrast are then given by Eqs.~(\ref{equ:MFrel2}), (\ref{eq:var}) and (\ref{eq:nphilin}), which to leading 
order in $M^2/H^2$ give
\be
n_\delta^{\rm lin-MF}-1=\frac{2}{3}\frac{m^2+3\lambda\varphi^2}{H^2},
\quad
A_\delta^{\rm lin-MF}
=4
\frac{\left(1+\frac{\lambda}{m^2}\varphi^2\right)^2}{\left(1+\frac{\lambda}{2m^2}\varphi^2\right)^2}
.\label{lin-MF}
\ee
In the quadratic limit, $\alpha\rightarrow \infty$ or $\lambda\rightarrow 0$,
this gives
\be
n_\delta^{\rm lin-MF}=1+\frac{2}{3}\frac{m^2}{H^2},\quad
A_\delta^{\rm lin-MF}=4.
\ee
In the quartic limit, $\alpha\rightarrow 0$ or $m\rightarrow 0$, %we obtain
%For example, the quartic potential then has the linear expansion
%\be
%V^{\rm lin}({\phi})=\f{\lambda}{4}\varphi^4+{\lambda}\varphi^3\delt%a{\phi}+\cdots\,,
%\ee
%giving %to leading order in $\hat{\phi}$ for the density contrast
%naively the density contrast
%\be
%{\delta}^{\rm lin}=4\f{\delta{\phi}}{\varphi}\,.
%\ee
%From Eq.~(\ref{eq:nphilin}), and using $m^2=3\lambda\varphi^2$,
one finds
\be
n_\delta^{\rm lin-MF}=%n_\phi^{\rm lin-MF}\approx
1+\frac{2\lambda\varphi^2}{H^2}
\approx 1+0.22508\lambda^{1/2}
%4-\sqrt{9-1.58117\lambda^{1/2}}
%\approx 
%1+0.26353\lambda^{1/2}
,
\quad A_\delta^{\rm lin-MF}=16,
\ee
where $\varphi$ was obtained from Eq.~(\ref{eq:var}) to leading order in $M^2/H^2$.
%via 
%\be\varphi^2=\f{3H^4}{8\pi^2(m^2+3\lambda\varphi^2)\,,}
%\ee
%which essentially is the Hartree-Fock approximation. 
As comparison with the correct values (\ref{eq:nquadratic0}) and (\ref{eq:nquartic}) shows, this is not a good approximation in either case.

\subsubsection{Stochastic-MF approximation}
\label{sec:3}
An alternative way of proceeding from Eq.~(\ref{eq:linrelation}) is to use the full stochastic field correlator {(\ref{eq:contphi})} for the right-hand side.
The relations (\ref{equ:MFrel2}) are still valid, so 
{using Eq.~(\ref{equ:quartfield})}
one obtains
\be
n_\delta^{\rm sto-MF}=n_\phi=1+2\frac{\Lambda_1}{H},
\quad
A_\delta^{\rm sto-MF}=\left(\frac{V'(\varphi)}{V(\varphi)}\right)^2\left(\phi_1\right)^2.
\label{eq:lin1}
\ee
In the limit $\alpha\rightarrow \infty$, Eq.~(\ref{eq:lin1}) gives
\be
n_\delta^{\rm sto-MF}=1 + \frac{2}{3}\frac{m^2}{H^2},
\quad
A_\delta^{\rm sto-MF}=4\f{(\phi_1)^2}{\varphi^2}=4,%=
 %4\bigg[\left(\frac{3}{2\pi^2}\right)^{1/2}{\frac{\Gamma(\frac{3}{4})}{\Gamma(\frac{1}{4})}} \bigg]^{-1/2}|(\tilde{\phi}^1)_1|
 %\approx 15.751\,,
\label{eq:lin2a}
\ee
and
in the limit $\alpha\rightarrow 0$, 
\be
n_\delta^{\rm sto-MF}=1 + 0.17785\lambda^{1/2},
\quad
A_\delta^{\rm sto-MF}=16\f{(\phi_1)^2}{\varphi^2}%=
 %4\bigg[\left(\frac{3}{2\pi^2}\right)^{1/2}{\frac{\Gamma(\frac{3}{4})}{\Gamma(\frac{1}{4})}} \bigg]^{-1/2}|(\tilde{\phi}^1)_1|
 \approx 15.751\,,
\label{eq:lin2}
\ee
from (\ref{hstar}) and (\ref{eq:quartic}).
Again, comparing with (\ref{eq:nquartic}), we see that this is a poor approximation.

\subsubsection{Linear-Gaussian approximation}
{Another way to express the density contrast correlator in terms of the field correlators is to assume that the field distribution is Gaussian.
Then there is no need to introduce a non-zero background field. Instead, the density contrast correlator
}
%Assuming the field distribution is Gaussian,  the correlator of the density contrast 
can be computed using Wick's theorem,
\begin{align}
&
{\langle\delta(0)\delta(\vec{x})\rangle^{\rm Gauss}=
\frac{\left\langle V(\phi(0)) V(\phi(\vec{x}))\right\rangle^{\rm Gauss}}{\left(\langle V(\phi)\rangle^{\rm Gauss}\right)^2} -1
}\nonumber\\&=
2\frac{
\langle \phi(0)\phi(\vec{x})\rangle^2
+6\frac{\lambda}{m^2}\langle \phi(0)\phi(\vec{x})\rangle^2
\langle \phi^2\rangle
+9\frac{\lambda^2}{m^4}\langle \phi(0)\phi(\vec{x})\rangle^2
\langle\phi^2\rangle^2
+3\frac{\lambda^2}{m^4}\langle \phi(0)\phi(\vec{x})\rangle^4
}
{\left(
\langle\phi^2\rangle+\frac{3\lambda}{2m^2}\langle\phi^2\rangle^2
\right)^2}
\nonumber\\
&\sim\frac{2}{\langle\phi^2\rangle^2}
\frac{\left(1+3\frac{\lambda}{m^2}\langle\phi^2\rangle\right)^2}
{\left(1+\frac{3}{2}\frac{\lambda}{m^2}\langle\phi^2\rangle\right)^2}
\langle \phi(0)\phi(\vec{x})\rangle^2,
\label{eq:Wick}
\end{align}
where in the last expression we have dropped the last term because it is subdominant at long distances.
From this we see that the Gaussian approximation implies an
exact relationship between the spectral indices of the field and the density contrast,
\be
n_\delta^{\rm Gauss}-1=2\left(n_\phi-1\right)
,
\label{eq:Gaussnrel}
\ee
which is correctly reproduced by the stochastic results (\ref{eq:nquadratic0}) in the free limit $\alpha\rightarrow\infty$, but {is not true} for general $\alpha$, as
Eqs.~(\ref{eq:nAgeneral0})--(\ref{eq:nAgeneral}) show.

In the linear-Gaussian approximation we make use of the linear result (\ref{eq:linearisedsol}) (with no mean field) and express the amplitude and the spectral index of the density contrast correlator
to leading order in $m^2/H^2$ as
\begin{align}
n_\delta^{\rm lin-Gauss}-1&\approx\frac{4}{3}\frac{m^2}{H^2},\nonumber \\ A_\delta^{\rm lin-Gauss}&=
2\frac{\left(1+3\frac{\lambda}{m^2}\langle\phi^2\rangle\right)^2}
{\left(1+\frac{3}{2}\frac{\lambda}{m^2}\langle\phi^2\rangle\right)^2}
\bigg(\f{A_\phi^{\rm lin}}{\langle\phi^2\rangle}\bigg)^2
\approx
2\frac{\left(1+\frac{9}{8\pi^2\alpha^2}\right)^2}{\left(1+\frac{9}{16\pi^2\alpha^2}\right)^2}
\rightarrow
\begin{dcases}
2
%\frac{2}{\langle\phi^2\rangle^2}
%\left(\frac{3}{8\pi^2}\frac{H^4}{m^2}\right)^2,
&\mbox{as $\alpha\rightarrow \infty$},
\\
8
%\frac{8}{\langle\phi^2\rangle^2}
%\left(\frac{3}{8\pi^2}\frac{H^4}{m^2}\right)^2,
&\mbox{as $\alpha \rightarrow 0$},
\end{dcases}
%\nonumber\\
%n_\delta^{\rm Gauss}-1&=&%4c\approx
%\frac{4}{3}\frac{m^2}{H^2},
\label{lin-Gauss}
\end{align}
where the spectral index is independent of $\alpha$.
Comparing with the full results (\ref{eq:nquadratic0}) and~(\ref{eq:nquartic}), we conclude that the linear-Gaussian approximation (unsurprisingly) works perfectly in the free case but fails when interactions are important.

\subsubsection{Stochastic-Gaussian approximation}
\label{sec:SG}
{As in Section~\ref{sec:3},
one can also}
%A better alternative to the linear approximation is
%to 
use the full stochastic field correlator (\ref{eq:contphi}) instead of the linearised solution (\ref{eq:linearisedsol}). If one still assumes a Gaussian field distribution, one can use Wick's theorem (\ref{eq:Wick}) to express the density contrast correlators in terms of it.

In this stochastic-Gaussian approximation the field amplitude $A_\phi$ and spectral index $n_\phi$ are, of course, the same as in the full stochastic case (\ref{eq:nAgeneral}). The spectral index and amplitude for the density contrast are given by Eq.~(\ref{eq:Gaussnrel}) and (\ref{eq:Wick}) %with the stochastic results for $n_\phi$ and $\langle\phi(0)\phi(\vec{x})\rangle$
\be
n_\delta^{\rm sto-Gauss}-1=4\frac{\Lambda_1}{H},\quad A_\delta^{\rm sto-Gauss}=2
\frac{\left(1+3\frac{\lambda}{m^2}\langle\phi^2\rangle\right)^2}
{\left(1+\frac{3}{2}\frac{\lambda}{m^2}\langle\phi^2\rangle\right)^2}
\frac{\left(\phi_1\right)^4}{\langle\phi^2\rangle^2}\,.
\label{eq:Ga1}
\ee
%\be
%A_\delta^{\rm Gauss}=2
%\frac{\left(1+3\frac{\lambda}{m^2}\langle\phi^2\rangle\right)^2}
%{\left(1+\frac{3}{2}\frac{\lambda}{m^2}\langle\phi^2\rangle\right)^2}
%\frac{\left(\phi_1\right)^4}{\langle\phi^2\rangle^2}.
%\label{eq:Ga2}
%\ee
In the free limit, $\alpha\rightarrow\infty$ or equivalently $\lambda\rightarrow 0$, this reproduces the full stochastic result (\ref{eq:nquadratic0}).
In the quartic limit, $\alpha\rightarrow 0$ or equivalently $m\rightarrow 0$, it gives
%
%The Gaussian approximation amounts to assuming that all correlators obey Wick's theorem. The density contrast for the quartic potential can then be calculated with the help of
%\be
%\langle{\phi}^4\rangle^{\rm Gauss}=3\langle{\phi}^2\rangle^2\,;\qquad\langle{\phi}^4(0){\phi}^4(\vec{x})\rangle^{\rm Gauss}=9\langle{\phi}^2\rangle^4+72\langle{\phi}(0){\phi}(\vec{x})\rangle^2\langle{\phi}^2\rangle^2+\cdots\,,
%\ee
%which straightforwardly gives {\color{red} for a quartic potential $V(\phi)=(\lambda/4)\phi^4$}
%\be
%\langle{\delta}(0){\delta}(\vec{x})\rangle^{\rm Gauss}
%=\frac{72\langle{\phi}(0){\phi}(\vec{x})\rangle^2\langle{\phi}^2\rangle^2}{\left(3\langle{\phi}^2\rangle^2\right)^2}
%\overset{|\vec{x}|H \gg1}{\approx}\label{eq:contG}
%\frac{8}{\langle{\phi}^2\rangle^2}\f{(\phi_1)^4}{(|\vec{x}| H)^{4\Lambda_1/H}}\,,
%\ee
%which gives 
%the parameters (\ref{equ:powerlaw})
\begin{align}
 n_\delta^{\rm sto-Gauss}\equiv 1+4\f{\Lambda_1}{H}\approx 1 + 0.35570\lambda^{1/2},\quad A_\delta^{\rm sto-Gauss}\equiv 8\f{(\phi_1)^4}{\langle{\phi}^2\rangle^2}%=\sqrt{8}\bigg[\left(\frac{3}{2\pi^2}\right)^{1/2}{\frac{\Gamma(\frac{3}{4})}{\Gamma(\frac{1}{4})}} \f{8\pi^2}{3}\bigg]^{-1}\big[(\tilde{\phi}^1)_1\big]^2
 \approx 7.753\,,
\end{align}
{which again fails to reproduce the full result (\ref{eq:nquartic}) but fares better than the other approximations}.

{In Fig.~\ref{fig:delta}, we can see that both mean field approximations, linear-MF and stochastic-MF, perform poorly at all values of $\alpha$. This is largely because the assumption of a non-zero mean field is not justified. The Gaussian approximations, linear-Gaussian and stochastic-Gaussian, work well at high $\alpha$, but then deviate from the full result {at} smaller $\alpha$, when the interactions become more important. Overall, the stochastic approximations tend to work slightly better than the linear ones. Therefore, out of all the approximation schemes we have presented, the stochastic-Gaussian approximation is the most accurate. However, even it cannot be considered to be a good approximation away from the non-interacting limit.
Therefore, the clear conclusion is that the full stochastic spectral expansion method is far superior to all of these approximation schemes.
}

%Out of all the approximation schemes we have presented the stochastic-Gaussian approximation fares the best, 
%however still with a {\color{red}significant} deviation from the full results 
%(\ref{eq:nquartic}) at the quartic limit.
\begin{figure}\begin{center}
		\includegraphics[width=0.97\textwidth,trim={0cm 0 0 0},clip]{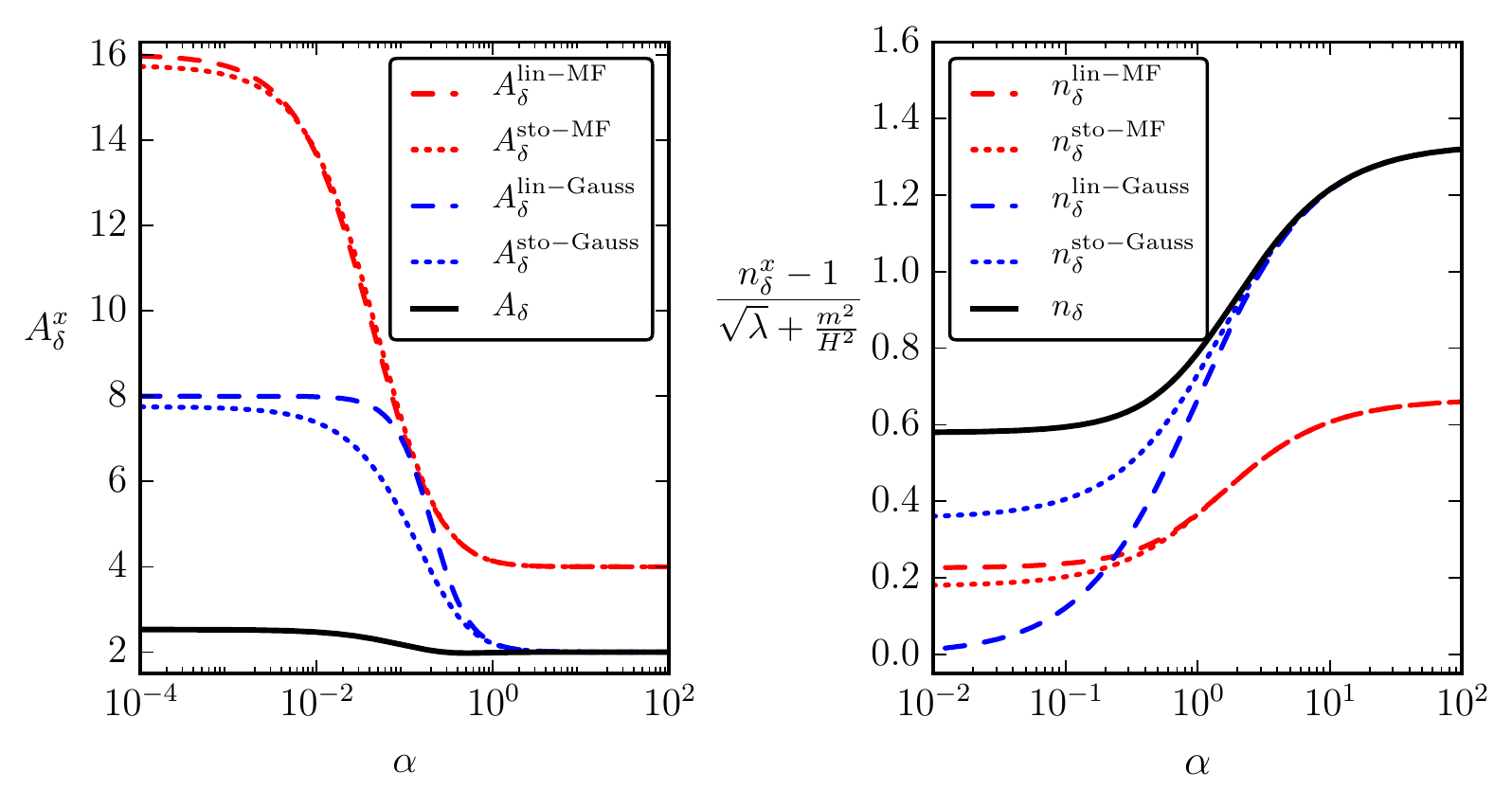}
	\end{center}
	\caption{\label{fig:delta}The full result for the leading spectral coefficient (left) and the spectral index (right) for the density contrast (\ref{eq:nAgeneral}) along with the approximations (\ref{lin-MF}), (\ref{eq:lin2}), (\ref{lin-Gauss}) and (\ref{eq:Ga1}). Note that $(n^x_\delta-1)/(\sqrt{\lambda}+\f{m^2}{H^2})$ is a function of only a single parameter, $\alpha$.}
\end{figure}

\section{Conclusions}
\label{sec:conclusions}

The stochastic {spectral expansion} is a very powerful tool for computing scalar correlation functions at long distances in de Sitter space. We have calculated them explicitly for a scalar field with a mass term and a self-interaction term, by solving the Schrödinger-like eigenvalue equation numerically to high precision. 
The resulting power spectra for the field and its density contrast are of the form (\ref{equ:Presults}) with the parameters given in Figs.~\ref{fig:eigenvalues} and~\ref{fig:fs}. We also considered the two limits of a free massive scalar ($\alpha\rightarrow\infty$) and a massless self-interacting scalar ($\alpha\rightarrow 0$) separately, with the power spectra given in Eqs.~(\ref{equ:Pfree}) and (\ref{equ:Pmassless}), respectively.

Comparison with different approximative schemes used in the literature shows that, except in the non-interacting limit, they fail to describe the correlators accurately at long distances. In particular, they {all} predict too high an amplitude and too low a spectral index for the density contrast. Therefore they tend to significantly overestimate its amplitude on large scales, which are relevant for any cosmological observations. This was found to be important, for example, in the case of spectator dark matter~\cite{Markkanen:2018gcw}, where the higher spectral index makes the isocurvature amplitude low enough to satisfy observational constraints.

%%%%%%%%%%%%%%%%%%%%%%%%%%%%%%%%%%%%%%%%%%%%%%%%%%%%%%%%%%%%%%%%%%%%%%%%%%
\section*{Acknowledgments}
We thank Marc Kamionkowski, Sami Nurmi and Gerasimos Rigopoulos for discussions. TM and AR are supported by the U.K. Science and Technology Facilities Council grant ST/P000762/1, and TM also by the Estonian Research Council via the Mobilitas Plus grant MOBJD323. TT is supported by the Simons foundation. SS is funded by Royal Society grant RGF\textbackslash EA\textbackslash 180172

%%%%%%%%%%%%%%%%%%%%%%%%%%%%%%%%%%%%%%%%%%%%%%%%%%%%%%%%
\bibliography{HiggsFluctuations}

%%%%%%%%%%%%%%%%%%%%%%%%%%%%%%%%%%%%%%%%%%%%%%%%%%%%%%%%%%%%%%%%%%%%%%%%%%%%%%%%%%%%%%%%%%

\end{document}